\documentclass[journal]{IEEEtran}

\usepackage{tkz-graph}
\usepackage{tikzscale}

\usepackage{cite}
\usepackage{amsmath,amssymb,amsfonts}
\usepackage{algorithmic}
\usepackage{graphicx}
\usepackage{textcomp}
\usepackage{xcolor}
\usepackage{enumitem} 
\def\BibTeX{{\rm B\kern-.05em{\sc i\kern-.025em b}\kern-.08em
    T\kern-.1667em\lower.7ex\hbox{E}\kern-.125emX}}
\begin{document}
\title{Enabling Deletion in Append-Only Blockchains 
(Short Summary / Work in Progress)
}

\author{
	\IEEEauthorblockN{Michael Kuperberg\\}
	\IEEEauthorblockA{
		\textit{Blockchain and Distributed Ledgers Group, DB Systel GmbH}\\ 
		Frankfurt am Main, Germany \\
		michael.kuperberg@deutschebahn.com
	}
}

\maketitle

\begin{abstract}
Conventional blockchain implementations with append-only semantics do not support deleting or overwriting data in confirmed blocks. 
However, many industry-relevant use cases require the ability to delete data, especially when personally identifiable information is stored or when data growth has to be constrained.
Existing attempts to reconcile these contradictions compromise on core qualities of the blockchain paradigm, as they include backdoor-like approaches such as central authorities with elevated rights or usage of specialized chameleon hash algorithms in chaining of the blocks. 
In this technical report, we outline
a novel architecture for the blockchain ledger and consensus, which uses a tree of context chains with simultaneous validity.
A context chain captures the transactions of a closed group of entities and persons, thus structuring blocks in a precisely defined way. 
The resulting context isolation enables consensus-steered deletion of an entire context without side effects to other contextes. 
This architecture opens the possibility of truncation, data rollover and separation of concerns, and can help to fulfill the GDPR regulations. 

\end{abstract}

\begin{IEEEkeywords}
DLT, 
distributed ledgers, 
blockchain, 
deletion, 
erasability, 
truncation, 
rollover, 
personally identifiable information, 
PII, 
GDPR, 
consensus protocols, 
WORM, 
append-only storage, 
decentralization
\end{IEEEkeywords}

\section{Introduction}
\label{Introduction}

Most blockchains and distributed ledgers follow an append-only, write-once read-many (WORM) approach to ensure auditability and trustworthiness. 
This pattern means that the inability to delete deters the adaption of these implementations, especially where law-mandated removal of person-related data is obligatory. 
Additionally, the append-only semantics imply that blockchain length and data size grow continously, which makes a long-term blockchain usage problematic. 
The latter concern is partially addressed in some scenarios by minimizing the amount of ledger-stored data: for example, some applications choose to only store hashes (fingerprints) of the data on-chain - yet such a restriction reduces the utility of a shared ledger. 
Consequently, the adoption of the distributed ledger technology (DLT) and specifically of blockchains is hindered by an inflexible approach to auditability. 

The mantra of ``data cannot be deleted from the blockchain'' is not a requirement that must be implemented at any cost. 
It is rather the consequence of cryptographic chaining of blocks, which itself is an implementation pattern chosen to fulfill the requirements of tamper-awareness (through integrity checks) and a decision made in the early blockchain implementations. 
In this report, we argue that deletion should be possible in a distributed ledger (blockchain), but it should be a consensus-based action and it must remove data in a way that does not compromise the integrity of other blocks and their data. 
The contribution of this 
report 
is therefore a tree-based structure to store the ledger, which groups transactions based on business contextes and into a linear subchain within the ledger tree where all subchains are valid at the same time. 
This design leads to an implementation where linear subchains can be deleted without affecting the other subchains, and such deletion is decided and agreed in a consensual way. 

The benefit of the proposed approach is that it maintains the important qualities of the blockchain paradigm (decentralization and tracability of decisions, through use of asymmetric cryptography to ensure the integrity of data, etc.) while relaxing the restrictions which are heavy obstacles to the applicability of the blockchains. 
It is important that this approach is different from forking
 (as e.g. in Ethereum) and also different from ``backdoor''-based approaches such as those relying on so-called chameleon hashes. 
Another advantage of the proposed solution is that it fits well into permissioned blockchains, but can also be used in permissionless solutions. 
From the business perspective, it supports such commonplace scenarios as year-end closings (where balances are computed so that detailed logs can be archived) and also general truncation/balancing. 
At the same time, the proposed approach does not allow the deletion (or overwriting) of arbitrary blocks and transactions, so that the auditability promise of the blockchain remains in place.  
%
%
The remainder of the report 
is structured as follows: 
Sec.~\ref{ContextChains} presents our contribution, 
Sec.~\ref{Evaluation} describes how we plan to evaluate our approach and
Sec.~\ref{Conclusions} concludes by outlining future work and current limitations of the presented approach. 
\section{Enabling Deletion in Blockchains and Distributed Ledgers With Chaining}
\label{ContextChains}
To explain our approach, consider a scenario of a conventional shared and decentralized, distributed linear ledger $L_{lin}$ with chained blocks of transactions.  
%
First, recall that a blockchain transaction can refer to one, two or more entities that can be PII (e.g. names or addresses). 
A simple statement about a person (when a blockchain would be used for identity management) may refer to only one PII entry. 
\begin{figure*} 
	[htbp]
	\begin{center}
		\includegraphics[trim = 0mm 0mm 0mm 0mm, clip, width=0.95\textwidth]
		{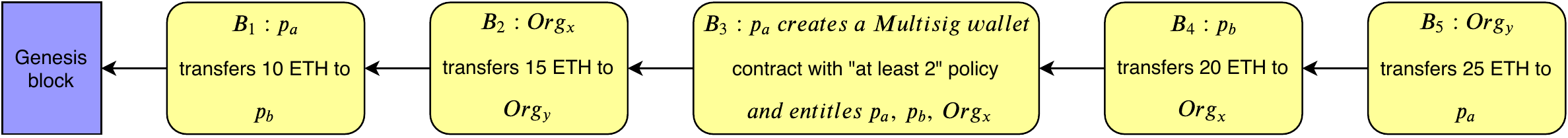}
		\caption{$L_{lin}$: conventional blockchain: transactions with different context scopes}
		\label{fig:linearityExample}
	\end{center}
\end{figure*}
But in Ethereum, for example, while a simple transaction can only send Ether from one address to one address, the \textit{definition} of an Ethereum 
smart contract (which is created through a special transaction) can refer to more than two \textit{hardcoded} Ethereum addresses of entities (for example, the definition of a contract that implements a multisignature wallet). 
Also, an Ethereum smart contract can have more than two input parameters which accept Ethereum addresses (or other types that can be of PII significance), with input values that change from invocation to invocation. 
Overall, a transaction in Ethereum can 
refer to more than two persons or organizations, and  
other 
blockchain/ledger implementations have similar properties. 

For explanation and illustration purposes, we start with the following initial simplifications: 
\begin{itemize}
	\item we assume that the blockchain network is operated in a permissioned way by a set of authorities (``organizations'', borrowing the terminology from Hyperledger Fabric)
	\item we start out with only two blockchain-participating organizations $Org_x$ and $Org_y$, and each of them runs at least one blockchain node with a replica of the ledger; both organizations participate in the consensus 
	\item we assume that the resulting blockchain contains transactions about only two physical persons $p_a$ and $p_b$ 
\end{itemize} 

In total, the transactions on this blockchain can include statements about any combination of $\{Org_x,Org_y,p_a,p_b\}$ including the empty set,  
for a max. of 
$2^4=16$ combinations. 

Fig.~\ref{fig:linearityExample} shows an example contents of $L_{lin}$ with above setup. 
For illustration purposes, each block contains only one transaction;
hashes and other block contents are not shown. 
Note how this linear chain mixes transactions that concern 
$p_a$ 
and/or 
$p_b$ 
with transactions which concern $Org_x$ and/or $Org_y$ or a mix of organization(s) and person(s). 
Clearly, if $p_a$ would request to have $p_a$'s data deleted from $L_{lin}$ (no matter whether an individual transaction(s) or 
all of them), this deletion would 
either invalidate $L_{lin}$ (by breaking the chaining if only some blocks/transaction are deleted), 
or require the entire $L_{lin}$ to be deleted as such, incl. transactions without reference to $p_a$. 

Even in products where a non-linear ledger structure is used (such as the Tangle \cite{iotatangle} DAG of IOTA), the remaining underlying problem is the \textit{unstructured} placement of transactions. 
The core solution aspect of the approach that we propose is a \textit{structured} placement of the transactions.  

\begin{figure*}
	[!t]
	\begin{center}
		\includegraphics[trim = 0mm 0mm 0mm 0mm, clip, width=0.98\textwidth]
		{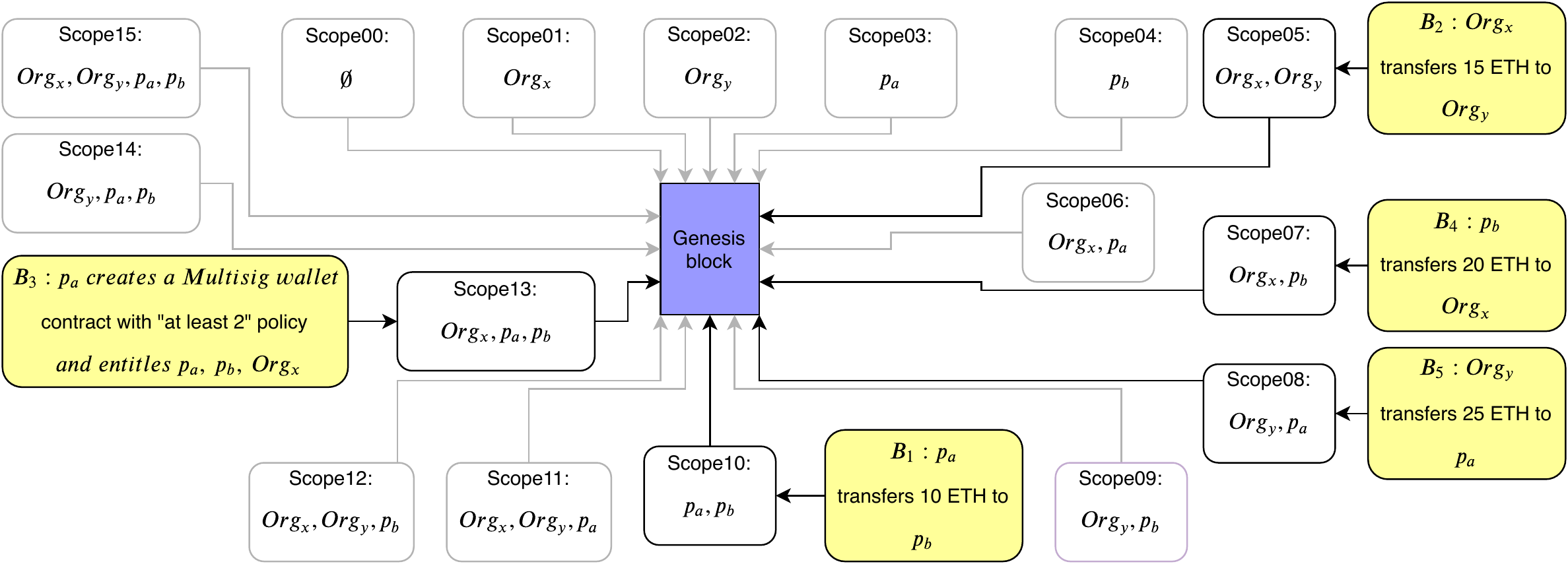}
		\caption{$L_{new}$: non-linear blockchain, transactions with the same context scope are located in the same branch (subtree); subtrees are arranged in clockwise manner; subroots shown in grey are not needed and should only be created on-demand once blocks with corresponding transactions are mined}
		\label{fig:newApproach}
	\end{center}
\end{figure*}
Thus, instead of linear chaining as in Fig.~\ref{fig:linearityExample}, consider our proposed new approach as illustrated by $L_{new}$ in Fig.~\ref{fig:newApproach}: anchored at the ``global'' root (usually containing a genesis block), we create a set of sub-chains with each having a sub-anchor (Scope00 through Scope15
for the above example
) which defines each subchain's scope. 
Continuing with the above example, the resulting DAG (directed acyclic graph) forms a tree with up to 16 well-defined branches - where all the branches are valid at the same time. 
In Fig.~\ref{fig:newApproach}, all transactions from Fig.~\ref{fig:linearityExample} have been placed into the appropriate sub-chain, following the scope descriptor in the sub-root. 
Transaction placement is based on \textit{precise} scope match. 
For example, a transaction with the scope $\{Org_x,p_a\}$ may only be placed under Scope06, but not under Scope11, Scope13 or Scope15. 

Note that for illustration purposes, all sub-roots in Fig.~\ref{fig:newApproach} have been created - even for branches where no transactions have been assigned to yet; 
to save space and computational efforts, creation of a sub-branch should be delayed up to the moment where a first transaction/block need to be added to that branch.
In general, the upper bound for the possible number of branches (given $k$ organizations and $m$ persons) is 
$$\sum_{i=0}^{(k+m)}\binom{(k+m)}{i}=2^{(k+m)}$$ 

We believe that only a small fraction of branches (well below the upper bound of $2^{(k+m)}$) indeed be created, and the storage/runtime overhead of the subroots is not a showstopper. 

Smart contracts (also stored on-chain) can be handled similarly to transactions that change the state of assets: smart contracts with source code whose declaration does reference PIIs (e.g. the identity/address of the smart contract owner in Ethereum) are stored in the corresponding context chain. 
Smart contracts whose source code do not contain any reference to PII can be stored in the context chain with the $\emptyset$ qualifier (see Scope00 in Fig.~\ref{fig:newApproach}). 
Note that it would be possible to arrange the roots of the context chains in several layers, rather than in one. 
However, this appears not to yield any space savings - but brings higher complexity as searching for the context root has to go beyond the ``first layer''. 

Now, if person $p_a$ requests \textit{all} of her data to be removed from $L_{new}$, this affects several branches;
we first discuss how the request-affected branches are identified and addressed. 
 
Branches with the scopes not including $p_a$ at all are not affected (there are 8 of them).
Removing branch with the trivial Scope03 ($\{p_a\}$) does not have side effects on any other branches, and does not concern the node-runners $Org_x$ or $Org_y$. 
However, for the remaining 16-8-1=7 branches, the deletion of transactions would affect other party/parties, and it is the consensus principle that is essential here: 
a transaction cannot be deleted unilaterily or in an unjustified way, i.e. against the rights and obligations of other concerned parties. 
For example, as the transaction in block $B_1$ describes $p_a$'s transfer to $p_b$, the blockchain operation should not permit $p_a$ to unilaterally have $B_1$ deleted, unless $p_b$ agrees. 

Therefore, a deletion consensus is required in general, for \textit{all} affected branches. 
For deleting the branches with the scopes $\{Org_x,p_a\}$ 
and 
$\{Org_y,p_a\}$, the consensus-finding appears rather straightforward: organizations $Org_x$ and $Org_y$ check if they have any rightful business interest to the transactions stored in those branches, and they can object to the deletion, preventing consensus. 

The consensus rules must thus exhibit the following ``all-must-agree'' behaviour:
\begin{enumerate}
	\item all organizations that are part of the scope have to be asked to agree and have to respond
	\item any organization that is part of the scope has the right to veto
	\item any organization that is part of the scope that does not respond (to the transaction/branch deletion request) is considered to have issued a veto
\end{enumerate}

However, neither $p_a$ nor $p_b$ are organizations, and they do not operate nodes in this example. 
Thus, it is up to the consensus-participating $Org_x$ and $Org_y$ to execute the consensus, even if they are not in the scope of the transactions. 
It is thus necessary to design a ``guardian mechanism'' which ensures that the rights of an in-scope person ($p_b$'s in the above example) are properly represented during deletion requests. 

In reality, blockchain-using applications tend not to permit all end users to act directly on blockchain. 
Instead, \textit{organizations} form blockchain business networks and transact together, representing people and application end users in varying ways. 
In particular, blockchain-using applications act \textit{on behalf of} the end user (``customer'' person) and the end user does not have insight into the blockchain-level data or code (smart contracts). 

In enterprise-grade blockchains such as Hyperledger Fabric, a person which is subject of a transaction (such as $p_a$ in the case of 
the transaction in $B_1$) is not required to be part of any organization, and can be represented in many ways, for example as UUIDs, multi-field identifiers, DIDs 
(see below) 
etc. 
- consequently, the same person can be represented through different identifiers. 
Even in blockchain implementations that do not have concepts of organizations and which are geared towards participating individuals (as it is the case in unpermissioned Ethereum), a transaction \textit{about} a person may not need involvement \textit{from} that person. 

We see different designs for identifying the potential ``endorsers'' of the deletion consensus, and for setting the ``endorsement policy''. 
For a given branch with the scope $S_g$, potential endorsers may be chosen based on these options:
\begin{enumerate}
	\item  all organizations that have endorsed the transactions that are included in $S_g$ (minus endorsers which have since left the blockchain network), or
	\item all organizations in the scope $S_g$, plus organizations which are named guardians of the persons in the scope $S_g$, 
	or
	\item all (any) organizations within the blockchain network
\end{enumerate}
Possible endorsing policy in every of these three cases should be ``all potential endorsers'', but depending on whether timeouts are supported in a given blockchain product and on how reliable the network is, a more flexible policy may be chosen, resulting in ``no response after timeout X means agreement'' implementation rather than in ``no response means veto''. 

Removing data from a context chain does not mean that the sub-root must be removed as well: it can remain in-place (keeping the scope definition in place) unless the requesting person insists on having her ``presence'' in the ledger removed as well (comparable to ``delete account''). 




It is clear that after a deletion has been agreed, each ledger-holding blockchain network participant must replicate the deletion on its side (including logs and backups),   
and DevOps/corporate security measures must be in place to prevent unauthorized copying of data. 

As stated above, a transaction is sorted into a context chain based on the declaration of its scope: a set of identifiers for persons and for organizations which are directly affected by that transaction. 
To ensure that the context is identically understood by all network participants and by the software on their nodes, we argue that the context should be made explicit by the transaction submitter, and should not have to be derived by the transaction validators, let alone by the ledger storage module. 
The identifiers to be used may vary depending on the specific ledger implementation, but we believe that the usage of DIDs (Decentralized Identifiers \cite{didspec2}, a W3C standard) may be beneficial when referencing humans or organizations. 

DIDs are primarilty used by Self-Sovereign Identity solutions (e.g. Sovrin \cite{sovrin} SSI), many of which use a ledger (e.g. Hyperledger Indy \cite{hyperledger-indy}) to store fingerprints (hashes) of Verifiable Credentials \cite{vc}. 
DIDs are used both for clients (identity holders) and service providers (incl. organizations). 
In SSI, the user remains in control over which data (incl. assertions and entitlements) is passed to which service provider - but the user is also solely responsible for safeguarding her identity information and doing backups. 
Most SSI offerings use a mobile device as biometrically-secured storage for identity information. 
As stated above, the identifiers within the context of the proposed context chain should be sorted for efficiency reasons; DIDs are simple three-section strings that can be sorted easily. 

Thus, the general ledger technology with our erasability feature and the SSI approach can converge from two sides: 
\begin{itemize}
	\item general ledger technology can refer to a represenation for persons and organizations that is vendor-independent, privacy-oriented and supported by an increasing number of tools (wallets, identity provider frameworks etc.)
	\item SSI technology can choose to store/exchange (enrypted) Verifiable Credentials on-ledger if deletion is enabled, since GDPR compliance is no longer out of reach
\end{itemize} 


\label{subSeveralTxPerBlock}

\section{Evaluation}
\label{Evaluation}

To implement the context chains and the consensus-based deletion, we have considered additions/modifications to the open source code of Hyperledger Fabric and Ethereum (geth). 
We believe that the presented concepts of context chains and deletion consensus are rather clear, even though a peer review of the concept is necessary before it can be refined and implemented. 
Likewise, we believe such an implementation should be preceded by a structured ``request for comments'' workflow, for example by creating an ERC (Ethereum Request for Comment) 
\cite{ercwebsite}. 
Therefore, this report and a detailed follow-up publication (under preparation) serve as the scientific foundation for such an ERC. 


From the efficiency perspective, the additional complexity and space requirements of our approach will have an impact on the performance of blockchain operations. 
However, we also believe that long-term benefit of erasability (especially lower storage demands) will outweigh the runtime overhead associated with context evaluation and with deletion consensus. 
In particular, if application architecture can store shared sensitive/PII data (which was previously confined to off-chain storage and off-ledger data distribution) on-chain, the overall complexity and resource overhead may decrease. 
This would work best in ledger which have mechanisms for selective data distributions (e.g. channels or the ``need-to-know'' principle). 

\section{Conclusions and Future Work}
\label{Conclusions}
In this report, we have presented a novel solution for adding deletion capabilities to append-only (WORM) blockchains: using the ``context chain'' architecture pattern, a separation of concerns leads to a non-linear ledger structure with an accompanying, clear rules of transaction placing. 
Context chains are complemented by consensus-driven decision making for deletion, ensuring that deletion is not endangering auditability and trustworthiness of the decentralized blockchain/ledger. 

We have discussed different aspects of the deletion problem, including ledgers that do not use chaining, non-cooperating (or absent) network participants and the effects of non-absolute majorities on the erasability 
of data. 
The opportunities that are unlocked by the ability to delete data in append-only blockchains include space savings, accordance with GDPR rules, and support of business processes such as rollover/balancing and truncation. 

For future work, we are already working on the
\begin{itemize}
	\item  analysis of sidechains and state channels (regarding deletion, pruning, GDPR compliance and branching)
	\item analysis of further attempts to provide erasability (e.g. through chameleon hashes \cite{ateniese2017a})
	\item applicability to non-linear ledgers such as 
	R3 Corda\cite{corda}
	\item truncation of suffixes and prefixes for a given ledger, as well as the special case of dangling (unconfirmed) blocks
	\item applicability of our concept to permissionless and organization-less ledgers, and to ledgers where a new block \textit{must} confirm more than one previous block
\end{itemize}

As part of our work, we plan pursue context chain implementation for a major enterprise-grade DLT implementation such as Hyperledger Fabric, R3 Corda 
or Quorum (permissioned Ethereum), 
since these open-source products do not provide facilities to delete transactions or blocks. 
Likewise, we plan to address erasability 
in ledgers which are used as foundations for self-sovereign identity, such as Sovrin's Plenum/Indy ledger (\cite{sovrin} / \cite{hyperledger-indy}). 
Finally, we plan to measure the performance and scalability of deletion, including the consensus phase. 
\section*{Acknowledgments}
Patrick Charrier 
and 
Michael Schnabel 
cross-validated the lack of deletion facilities in Hyperledger Fabric and in Ethereum (geth). 
Moritz von Bonin, 
Sviatoslav Butskyi,
Koraltan Kaynak, 
Robin Klemens, 
Artur Philipp,
and 
Malte Plescher
provided valuable feedback and suggested improvements to the text. 
Manfred Schulze 
provided much-valued support in patent research and drafting when the original idea presented in this report was filed in 2018. 

\bibliographystyle{IEEEtran}
\bibliography{Kuperberg2020c-Delete-On-Blockchain}

\end{document}